\documentclass[aps,pra,showpacs,twocolumn,amsmath,amssymb]{revtex4}
\usepackage{subfigure}
\usepackage{latexsym}
\usepackage{graphicx}
\usepackage{bm}

\graphicspath{{./figures/}}

\newcommand{\ddt}[1]{\frac{\partial #1}{\partial t}}

\newcommand{\dddtdt}[1]{\frac{\partial^2 #1}{\partial t^2}}

\newcommand{\ddz}[1]{\frac{\partial #1}{\partial z}}
\newcommand{\dddzdz}[1]{\frac{\partial^2 #1}{\partial z^2}}

\newcommand{\ket}[1]{\vert #1 \rangle}
\newcommand{\abs}[1]{\vert #1 \vert}

\begin{document}

%%% Title of article %%%
\title{Trapping of light pulses in ensembles of stationary $\Lambda$ atoms}

%%% Authors and affiliations %%%
\author{Kristian Rymann Hansen}
\author{Klaus M\o lmer}
\affiliation{Lundbeck Foundation Theoretical Center for Quantum
System Research, Department of Physics and Astronomy, University of
Aarhus, DK-8000 \AA rhus C, Denmark}

%%% Abstract %%%
\begin{abstract}
We present a detailed theoretical description of the generation of
stationary light pulses by standing wave electromagnetically induced
transparency in media comprised of stationary atoms. We show that,
contrary to thermal gas media, the achievable storage times are
limited only by the ground state dephasing rate of the atoms, making
such media ideally suited for nonlinear optical interactions between
stored pulses. Furthermore, we find significant quantitative and
qualitative differences between the two types of media, which are
important for quantum information processing schemes involving
stationary light pulses.
\end{abstract}
 \maketitle

%%% Beginning of main matter %%%
\section{Introduction}
The coherent transfer of quantum states between light and atoms has
been the subject of much research, both experimentally and
theoretically, motivated by potential applications in quantum
computing, quantum cryptography and teleportation. While the
transfer of quantum states from light to a single atom can in
principle be achieved by cavity QED techniques \cite{Cirac}, the
required strong-coupling regime is experimentally very difficult to
reach. To overcome this difficulty the use of atomic ensembles,
rather than single atoms, has been proposed
\cite{Kozhekin,Fleischhauer1,Kuzmich,Sherson} and implemented for
storage of classical light pulses \cite{Liu,Phillips}, and recently
storage of non-classical pulses has been demonstrated
\cite{Julsgaard,Eisaman}. One such light storage scheme is based on
electromagnetically induced transparency (EIT) \cite{Harris} in
ensembles of $\Lambda$-type atoms \cite{Fleischhauer1}. While the
storage and retrieval of light pulses with this scheme has been
demonstrated utilizing many different types of atomic ensembles,
including Bose-Einstein condensates \cite{Liu} and thermal gasses
\cite{Phillips} as well as solid state media
\cite{Turukhin,Longdell}, the non-trivial manipulation of, and
interaction between, stored pulses is hampered by the inherent
trade-off between storage time and field amplitude. A step towards
overcoming this problem was taken with the suggestion of using
standing wave fields to create a periodic modulation of either the
dispersive \cite{Andre1} or the absorptive \cite{Bajcsy} properties
of the medium, inducing a photonic bandgap \cite{Yariv} and creating
a stationary light pulse. Schemes to implement a controlled phase
gate using these techniques have been proposed using either the
dispersive \cite{Friedler} or the absorptive \cite{Andre2} grating
technique, but both are still hampered by a trade-off between
storage time and field amplitude. For the latter case, only thermal
gas media have been considered, and a detailed theoretical treatment
of this case is given in \cite{Zimmer}. This theory, however, does
not apply to media comprised of stationary atoms such as ultra cold
gasses or solid state media \cite{Hansen}. In this article we
present a detailed theoretical treatment of the creation of
stationary light pulses by the absorptive grating technique for
media comprised of stationary atoms. We find that the loss of
excitations inherent to the thermal gas case is absent for
stationary atoms, making such media ideally suited for the kind of
non-linear optical interactions envisaged in \cite{Andre2}.
Furthermore we find interesting quantitative and qualitative
differences between the thermal gas and ultra cold gas cases when
quasi-standing wave coupling fields are considered. These
differences are important for the proposed controlled phase gate
scheme \cite{Andre2} in stationary atom media.

In Sec.~\ref{Sec:Standing_wave_polaritons_cold} we present a
detailed account of our theory of stationary light pulses in media
comprised of stationary atoms and compare the results to the thermal
gas case. The theory is complemented in Sec.~\ref{Sec:Non-adiabatic}
by a calculation of non-adiabatic corrections. A summary of our
results is provided in Sec.~\ref{Sec:Summary}. Appendix
\ref{Sec:Standing_wave_polaritons_thermal} contains a brief review
of the theory of stationary light pulses in thermal gas media
\cite{Zimmer}, reformulated in terms of polariton fields, used for
comparison with the stationary atom case.

\section{Standing wave polaritons in ensembles of stationary atoms}
\label{Sec:Standing_wave_polaritons_cold}

We consider an ensemble of $N$ non-moving $\Lambda$ atoms
interacting with probe and coupling lasers propagating parallel to
the $z$ axis. The two lower states $\ket{b}$ and $\ket{c}$ of the
atoms (see Fig.~) are assumed to be nearly degenerate, such that the
magnitude of the wave vectors of the probe and coupling lasers can
be considered identical $(k_p\simeq k_c=k)$.
\begin{figure}
\includegraphics{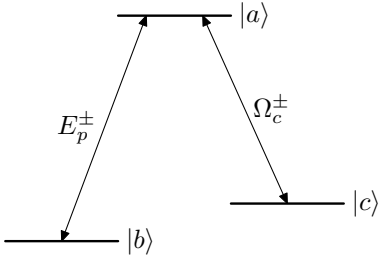}
\caption{\label{Fig:level_diagram}The 3-level $\Lambda$ atom. The
quantized probe field couples the ground state $\ket{b}$ to the
excited state $\ket{a}$, while the classical coupling field couples
the metastable state $\ket{c}$ to $\ket{a}$.}
\end{figure}

The Hamiltonian for the $N$ atom problem is
\begin{equation}
\hat{H}=\hat{H}_F+\sum_{j=1}^N\left(\hat{H}_A^j+\hat{H}_L^j
+\hat{H}_V^j\right)
\end{equation}
where $\hat{H}_F$ and $\hat{H}_A$ describe the free electromagnetic
field and the atoms, $\hat{H}_L$ describes the interaction of the
atoms with the probe and coupling fields, and $\hat{H}_V$ describes
the interaction with the vacuum field modes. The individual terms
are given by
\begin{subequations}
\begin{align}
\hat{H}_F&=\sum_m\hbar\omega_m\hat{a}_m^\dag\hat{a}_m \\
\hat{H}_A^j&=\hbar\omega_{cb}\hat{\sigma}_{cc}^j+\hbar\omega_{ab}
\hat{\sigma}_{aa}^j \\
\hat{H}_L^j&=-\bigl(\hat{\mathbf{E}}_p+\hat{\mathbf{E}}_c\bigr)\cdot
\bigl(\mathbf{d}_{ba}\hat{\sigma}_{ba}^j+\mathbf{d}_{ca}
\hat{\sigma}_{ca}^j+\mathrm{h.a.}\bigr) \\
\hat{H}_V^j&=-\hat{\mathbf{E}}_V\cdot\bigl(\mathbf{d}_{ba}\hat{\sigma}_{ba}^j
+\mathbf{d}_{ca}\hat{\sigma}_{ca}^j+\mathrm{h.a.}\bigr)
\end{align}
\end{subequations}
We introduce slowly varying field operators for the electromagnetic
field. Since we are allowing for standing wave fields, we write the
field operator as a superposition of two traveling wave fields
propagating in opposite directions
\begin{equation}
\hat{\mathbf{E}}_{p,c}(z,t)=\sqrt{\frac{\hbar\omega_{p,c}}{2\varepsilon_0V}}
\mathbf{e}_{p,c}E_{p,c}(z,t)e^{-i\omega_{p,c}t}+\text{h.c.}
\end{equation}
where the operators $E_{p,c}$ are given by
\begin{equation}\label{Eq:standingwave_field_decomp2}
E_{p,c}(z,t)=E_{p,c}^+(z,t)e^{ikz}+E_{p,c}^-(z,t)e^{-ikz}.
\end{equation}
The field operators $E_{p,c}^\pm$ are the slowly varying field
operators for the forward and backward propagating components of the
probe and coupling fields with carrier frequencies $\omega_{p,c}$,
and $\mathbf{e}_{p,c}$ are the respective polarization vectors.

We define continuum atomic operators $\hat{\sigma}_{\mu\nu}$ by
summing over the individual atoms in a small volume $V$, and
introduce slowly varying atomic operators $\sigma_{\mu\nu}$ defined
by
\begin{subequations}\label{Eq:SV_atom_opr_stand}
\begin{align}
%\hat{\sigma}_{bb}&=\sigma_{bb} \\
%\hat{\sigma}_{cc}&=\sigma_{cc} \\
%\hat{\sigma}_{aa}&=\sigma_{aa} \\
\hat{\sigma}_{ba}&=\sigma_{ba}e^{-i\omega_pt} \\
\hat{\sigma}_{ca}&=\sigma_{ca}e^{-i\omega_ct} \\
\hat{\sigma}_{bc}&=\sigma_{bc}e^{-i\left(\omega_p-\omega_c\right)t}.
\end{align}
\end{subequations}
Notice that the operators defined by (\ref{Eq:SV_atom_opr_stand})
are slowly varying in \emph{time}, but not in \emph{space}.

The Heisenberg-Langevin equations for these operators in the
rotating wave approximation are
\begin{subequations}
\begin{align}
\dot{\sigma}_{aa}&=-i\bigl(g_pE_p^\dag\sigma_{ba}+\Omega_c^*\sigma_{ca}
-\mathrm{h.a.}\bigr)-\gamma\sigma_{aa}+F_{aa} \\
\dot{\sigma}_{bb}&=i\bigl(g_pE_p^\dag\sigma_{ba}-\mathrm{h.a.}\bigr)
+\gamma_b\sigma_{aa}+F_{bb} \\
\dot{\sigma}_{cc}&=i\bigl(\Omega_c^*\sigma_{ca}-\mathrm{h.a.}\bigr)
+\gamma_c\sigma_{aa}+F_{cc} \\
\dot{\sigma}_{ba}&=i\bigl(g_pE_p(\sigma_{bb}-\sigma_{aa})+\Omega_c
\sigma_{bc}\bigr)-\Gamma_{ba}\sigma_{ba}+F_{ba} \\
\dot{\sigma}_{ca}&=i\bigl(\Omega_c(\sigma_{cc}-\sigma_{aa})+g_pE_p
\sigma_{bc}^\dag\bigr)-\Gamma_{ca}\sigma_{ca}+F_{ca} \\
\dot{\sigma}_{bc}&=i\bigl(\Omega_c^*\sigma_{ba}-g_pE_p\sigma_{ca}^\dag
\bigr)-\Gamma_{bc}\sigma_{bc}+F_{bc}
\end{align}
\end{subequations}
where $\gamma=\gamma_b+\gamma_c$ is the decay rate of the excited
state $\ket{a}$ into the two lower states. The complex decay rates
$\Gamma_{\mu\nu}$ are given by
\begin{align}
\Gamma_{ba}&=\gamma_{ba}-i\delta_p, \\
\Gamma_{ca}&=\gamma_{ca}-i\delta_c, \\
\Gamma_{bc}&=\gamma_{bc}-i\Delta,
\end{align}
where $\gamma_{\mu\nu}$ are the dephasing rates of the respective
coherences, $\delta_{p,c}$ are the one-photon detunings of the probe
and coupling lasers, respectively, and $\Delta$ is the two-photon
detuning. We have also assumed that the coupling field can be
treated as a classical field with Rabi frequency $\Omega_c$ given by
\begin{equation}
\Omega_c(z,t)=\Omega_c^+(z,t)e^{ikz}+\Omega_c^-(z,t)e^{-ikz}.
\end{equation}
In the following we shall disregard the noise operators $F_{\mu\nu}$
since we will be considering the adiabatic limit.

\subsection{Weak probe approximation}
In order to solve the propagation problem, we assume that the probe
field is weak compared to the coupling field and that the probe
photon density is small compared to the atomic density. In this case
the Heisenberg-Langevin equations can be solved perturbatively. To
first order in the probe field amplitude, the relevant
Heisenberg-Langevin equations are
\begin{subequations}\label{Eq:weak_probe_HL_stand}
\begin{align}
\sigma_{ba}&=\frac{1}{i\Omega_c^*}\left(\Gamma_{bc}+\ddt{}\right)
\sigma_{bc}\label{Eq:weak_probe_HL_stand_a} \\
\sigma_{bc}&=-\frac{g_pE_p}{\Omega_c}-\frac{i}{\Omega_c}\left(
\Gamma_{ba}+\ddt{}\right)\sigma_{ba}
\end{align}
\end{subequations}
Combining equations (\ref{Eq:weak_probe_HL_stand}) we can obtain a
differential equation for $\sigma_{bc}$
\begin{equation}\label{Eq:diffeqn_sigmabc}
\sigma_{bc}=-\frac{g_pE_p}{\Omega_c}-\frac{1}{\Omega_c}\left(
\Gamma_{ba}+\ddt{}\right)\left[\frac{1}{\Omega_c^*}\left(
\Gamma_{bc}+\ddt{}\right)\sigma_{bc}\right]
\end{equation}

\subsection{Adiabatic limit}
In order to solve equation (\ref{Eq:diffeqn_sigmabc}), we consider
the adiabatic limit in which the fields vary slowly in time.
Introducing a characteristic timescale $T$ of the slowly varying
operators, we expand $\sigma_{bc}$ in powers of
$(\gamma_{ba}T)^{-1}$. To zeroth order we find
\begin{equation}\label{Eq:sigma_bc_stand_adia}
\sigma_{bc}=-\frac{g_pE_p}{\Omega_c}.
\end{equation}
Inserting this expression into (\ref{Eq:weak_probe_HL_stand_a}), we
find an expression for $\sigma_{ba}$ valid in the adiabatic limit
\begin{equation}\label{Eq:sigma_ba_stand_adia}
\sigma_{ba}=-\frac{1}{i\Omega_c^*}\left(\Gamma_{bc}+\ddt{}\right)
\left(\frac{g_pE_p}{\Omega_c}\right).
\end{equation}
By inserting the field decomposition
(\ref{Eq:standingwave_field_decomp2}) into the adiabatic expression
for $\sigma_{ba}$ (\ref{Eq:sigma_ba_stand_adia}) we get
\begin{equation}\label{Eq:sigma_ba_stand_adia2}
\begin{split}
\sigma_{ba}&=\frac{-g_p\left(\Gamma_{bc}+\ddt{}\right)}{i\Omega
(1+2\abs{\kappa^+}\abs{\kappa^-}\cos(2kz+\phi))} \\
&\quad\times\left(\frac{E_p^+e^{ikz}+E_p^-e^{-ikz}}{\Omega}\right),
\end{split}
\end{equation}
where we have also introduced the time dependent total Rabi
frequency $\Omega(t)=\sqrt{\abs{\Omega_c^+}^2+\abs{\Omega_c^-}^2}$
and the ratios $\kappa^\pm=\frac{\Omega_c^\pm}{\Omega}$ which are
assumed to be constant. The phase angle $\phi$ is defined by the
relation
\begin{equation}
\kappa^+\kappa^{-*}=\abs{\kappa^+}\abs{\kappa^-}e^{i\phi}.
\end{equation}

\subsection{Polariton field}
We now introduce a dark-state polariton (DSP) field analogous to the
DSP field defined in \cite{Fleischhauer1}
\begin{equation}\label{Eq:def_polariton_stand}
E_p^\pm(z,t)=\cos\theta(t)\Psi^\pm(z,t),
\end{equation}
where the angle $\theta$ is given by the total coupling laser Rabi
frequency through
\begin{equation}
\tan\theta(t)=\frac{g_p\sqrt{N_\mathbf{r}}}{\Omega(t)}.
\end{equation}
By inserting the definition (\ref{Eq:def_polariton_stand}) of the
DSP field into (\ref{Eq:sigma_ba_stand_adia2}) we get
\begin{equation}\label{Eq:sigma_ba_stand_adia3}
\begin{split}
\sqrt{N_\mathbf{r}}\sigma_{ba}&=\frac{-\bigl(\Gamma_{bc}+\ddt{}
\bigr)}{i\Omega\bigl(1+2\abs{\kappa^+}\abs{\kappa^-}\cos(2kz
+\phi)\bigr)}\\
&\quad\times\Bigl(\sin\theta\bigl(\Psi^+e^{ikz}+\Psi^-e^{-ikz}\bigr)\Bigr).
\end{split}
\end{equation}
To derive wave equations for the components of the DSP field, we
need to expand the optical coherence $\sigma_{ba}$ in spatial
Fourier components. We do this by inserting the Fourier series
\begin{equation}\label{Eq:Fourier_series1}
\frac{1}{1+y\cos x}=\frac{a_0}{2}+\sum_{n=1}^\infty a_n\cos(nx),
\end{equation}
where $y=2\abs{\kappa^+}\abs{\kappa^-}$ and $x=2kz+\phi$, into
(\ref{Eq:sigma_ba_stand_adia3}). Note that $y\leq 1$ which
guarantees the existence of the Fourier series except in the case of
a standing wave coupling field $(y=1)$. Fortunately, we can treat
this case successfully by considering the limit $y\rightarrow 1$ at
the end of our calculation.

Inserting the Fourier series into (\ref{Eq:sigma_ba_stand_adia3}) we
find
\begin{equation}\label{Eq:sigma_ba_Fourier}
\begin{split}
\sqrt{N_\mathbf{r}}\sigma_{ba}&=\frac{i}{\Omega}\left(\frac{a_0}{2}+
\sum_{n=1}^\infty \frac{a_n}{2} \left( e^{in(2kz+\phi)}
+e^{-in(2kz+\phi)}\right)\right) \\
&\quad\times\left(\Gamma_{bc}+\ddt{}\right)
\left(\sin\theta\left(\Psi^+e^{ikz}+\Psi^-e^{-ikz}\right)\right).
\end{split}
\end{equation}
From (\ref{Eq:sigma_ba_Fourier}) we see that $\sigma_{ba}$ can we
written as
\begin{equation}\label{Eq:sigma_ba_expansion}
\sigma_{ba}=\sum_{n=-\infty}^\infty \sigma_{ba}^{(2n+1)}
e^{i(2n+1)kz}.
\end{equation}
To derive a set of wave equations for the polariton field
components, we need to calculate the components $\sigma_{ba}^{+1}$
and $\sigma_{ba}^{-1}$ of the expansion
(\ref{Eq:sigma_ba_expansion}) which we label $\sigma_{ba}^\pm$ for
brevity. These components are given by
\begin{subequations}\label{Eq:sigma_ba_Fourier2}
\begin{align}
\sqrt{N_\mathbf{r}}\sigma_{ba}^+&=\frac{i}{2\Omega}\left(\Gamma_{bc}
+\ddt{}\right)[\sin\theta(a_0\Psi^++a_1e^{i\phi}\Psi^-)] \\
\sqrt{N_\mathbf{r}}\sigma_{ba}^-&=\frac{i}{2\Omega}\left(\Gamma_{bc}
+\ddt{}\right)[\sin\theta(a_0\Psi^-+a_1e^{-i\phi}\Psi^+)].
\end{align}
\end{subequations}
We see that we only need to calculate the first two Fourier
coefficients $a_0$ and $a_1$ of the expansion
(\ref{Eq:sigma_ba_Fourier}). These are given by
\begin{subequations}\label{Eq:Fouriercoef_a}
\begin{align}
a_0&=\frac{1}{\pi}\int_{-\pi}^\pi \frac{\mathrm{d}x}{1+y\cos x}
=\frac{2}{\sqrt{1-y^2}} \\
a_1&=\frac{1}{\pi}\int_{-\pi}^\pi \frac{\cos x\mathrm{d}x}{1+y\cos
x} =2\frac{\sqrt{1-y^2}-1}{y\sqrt{1-y^2}}.
\end{align}
\end{subequations}

Inserting the adiabatic expression (\ref{Eq:sigma_ba_Fourier2}) for
$\sigma_{ba}^\pm$ into the wave equations for the probe field
components
\begin{subequations}\label{Eq:probe_waveeqn}
\begin{align}
\left(\ddt{}+c\ddz{}\right)E_p^+(z,t)&=ig_pN_\mathbf{r}\sigma_{ba}^+(z,t)\\
\left(\ddt{}-c\ddz{}\right)E_p^-(z,t)&=ig_pN_\mathbf{r}\sigma_{ba}^-(z,t),
\end{align}
\end{subequations}
we obtain a set of coupled wave equations for the DSP field
components
\begin{widetext}
\begin{subequations}\label{Eq:Polariton_waveeqn_cold1}
\begin{align}
\ddt{\Psi^+}+c\cos^2\theta'\ddz{\Psi^+}&=-\sin^2\theta'\biggl[
\Gamma_{bc}\Psi^++se^{i\phi}\biggl(\Gamma_{bc}+\ddt{}\biggr)\Psi^-
-s\dot{\theta}\frac{\cos\theta}{\sin\theta}\Bigl(y\Psi^+-e^{i\phi}
\Psi^-\Bigr)\biggr]\\
\ddt{\Psi^-}-c\cos^2\theta'\ddz{\Psi^-}&=-\sin^2\theta'\biggl[
\Gamma_{bc}\Psi^-+se^{-i\phi}\biggl(\Gamma_{bc}+\ddt{}\biggr)\Psi^+
-s\dot{\theta}\frac{\cos\theta}{\sin\theta}\Bigl(y\Psi^--e^{-i\phi}
\Psi^+\Bigr)\biggr],
\end{align}
\end{subequations}
\end{widetext}
where we have introduced a new angle $\theta'$ defined by
\begin{equation}
\tan\theta'=\sqrt{\frac{a_0}{2}}\frac{g_p\sqrt{N_\mathbf{r}}}{\Omega}
=\sqrt{\frac{a_0}{2}}\tan\theta,
\end{equation}
as well as the constant
\begin{equation}
s=\frac{a_1}{a_0}=\frac{\sqrt{1-y^2}-1}{y}.
\end{equation}
Since we are considering the adiabatic limit in which the coupling
field Rabi frequency changes slowly in time, we shall neglect the
last term on the rhs.~of (\ref{Eq:Polariton_waveeqn_cold1}) in the
following.

\subsection{Low group velocity limit}
In the experimentally relevant \emph{low group velocity limit}
$\cos^2\theta\ll 1$, the wave equations
(\ref{Eq:Polariton_waveeqn_cold1}) take the simpler form
\begin{subequations}\label{Eq:polariton_waveeqn2a}
\begin{align}
\left(\Gamma_{bc}+\ddt{}\right)\Psi^+ +\abs{\kappa^+}^2v_g
\ddz{\Psi^+}&=\kappa^+\kappa^{-*}v_g\ddz{\Psi^-}, \\
\left(\Gamma_{bc}+\ddt{}\right)\Psi^- -\abs{\kappa^+}^2v_g
\ddz{\Psi^-}&=-\kappa^{+*}\kappa^-v_g\ddz{\Psi^+},
\end{align}
\end{subequations}
in the case where $\abs{\kappa^+}\geq\abs{\kappa^-}$. In the
opposite case, $\abs{\kappa^+}\leq\abs{\kappa^-}$, the wave
equations are
\begin{subequations}\label{Eq:polariton_waveeqn2b}
\begin{align}
\left(\Gamma_{bc}+\ddt{}\right)\Psi^+ +\abs{\kappa^-}^2v_g
\ddz{\Psi^+}&=\kappa^+\kappa^{-*}v_g\ddz{\Psi^-}, \\
\left(\Gamma_{bc}+\ddt{}\right)\Psi^- -\abs{\kappa^-}^2v_g
\ddz{\Psi^-}&=-\kappa^{+*}\kappa^-v_g\ddz{\Psi^+}.
\end{align}
\end{subequations}
We have introduced the group velocity $v_g=c\cos^2\theta$ in
equations (\ref{Eq:polariton_waveeqn2a}) and
(\ref{Eq:polariton_waveeqn2b}), and have also made use of the fact
that in the low group velocity limit, $\cos^2\theta'\simeq
\sqrt{1-y^2}\cos^2\theta$.

\subsection{Initial conditions}
We shall consider the same kind of experiment as in \cite{Bajcsy} in
which a probe pulse, propagating under the influence of a
copropagating \emph{traveling wave} coupling field, is stored in the
medium and subsequently retrieved by a \emph{standing wave} coupling
field with $\abs{\kappa^+}\geq\abs{\kappa^-}$.

Assuming that the standing wave coupling field is switched on at
$t=0$, we need to find the initial conditions for the two components
of the DSP field $\Psi^\pm(z,0)$. The initial condition for the
Raman coherence is% given by
\begin{equation}\label{Eq:sigma_bc_init_cond_stand}
\sqrt{N_\mathbf{r}}\sigma_{bc}(z,0)=-\Psi(z,0),
\end{equation}
where $\Psi(z,0)$ is a known function of $z$ determined by the DSP
field prior to switching on the standing wave coupling field. Using
(\ref{Eq:sigma_bc_stand_adia}) and the definition of the DSP field
in the standing wave case (\ref{Eq:def_polariton_stand}), along with
the initial condition (\ref{Eq:sigma_bc_init_cond_stand}), we get
\begin{equation}
\Psi(z,0)\left(\kappa^+e^{ikz}+\kappa^-e^{-ikz}\right)=\Psi^+(z,0)
e^{ikz}+\Psi^-(z,0)e^{-ikz}.
\end{equation}
From this expression we see that the initial conditions for the
components of the DSP field $\Psi^\pm(z,0)$ are
\begin{equation}\label{Eq:Psi_pm_init_cond}
\Psi^+(z,0)=\kappa^+\Psi(z,0),\qquad\Psi^-(z,0)=\kappa^-\Psi(z,0).
\end{equation}
With the initial conditions (\ref{Eq:Psi_pm_init_cond}), we find the
solution
\begin{widetext}
\begin{subequations}\label{Eq:DSP_stand_sol2}
\begin{align}
\Psi^+(z,t)&=\frac{\kappa^+}{2}\Biggl[\left(1+
\frac{\beta}{\abs{\kappa^+}^2}\right)\Psi(z-\beta r(t),0)+\left(1-
\frac{\beta}{\abs{\kappa^+}^2}\right)
\Psi(z+\beta r(t),0)\Biggr]e^{-\Gamma_{bc}t}, \\
\Psi^-(z,t)&=\frac{\kappa^-}{2}\Biggl[\Psi(z-\beta
r(t),0)+\Psi(z+\beta r(t),0) \Biggr]e^{-\Gamma_{bc}t},
\end{align}
\end{subequations}
where
$\beta=\sqrt{\abs{\kappa^+}^2(\abs{\kappa^+}^2-\abs{\kappa^-}^2)}$
and $r(t)=\int_0^t c\cos^2\theta(t')\mathrm{d}t'$.
\end{widetext}

\subsection{Probe retrieval by a standing wave coupling field}
To determine the solution for a standing wave coupling field, we let
$\kappa^\pm\rightarrow\frac{1}{\sqrt{2}}$ in the solution
(\ref{Eq:DSP_stand_sol2}). In this limit we find the solution
\begin{subequations}
\begin{align}
\Psi^+(z,t)&=\frac{1}{\sqrt{2}}\Psi(z,0)e^{-\Gamma_{bc}t}, \\
\Psi^-(z,t)&=\frac{1}{\sqrt{2}}\Psi(z,0)e^{-\Gamma_{bc}t}.
\end{align}
\end{subequations}
%Contrary to the case of a thermal gas medium considered in
%\cite{Zimmer}, we see that no diffusive broadening of the pulse
%envelope is present in the ultra cold gas case.

As an example, we take the initial condition for the DSP field to be
$\Psi(z,0)=\Psi_0\exp(-(z/L_p)^2)$, where $L_p$ is the
characteristic length of the stored probe pulse. The polariton
amplitude $\Psi_0$ is related to the initial probe field amplitude
$E_0$ by $\Psi_0=E_0/\cos\theta_0$, where $\theta_0$ is determined
by the Rabi frequency of the traveling wave coupling field prior to
storage.

\begin{figure}
\subfigure[\ Stationary atoms]
{\includegraphics[width=4cm]{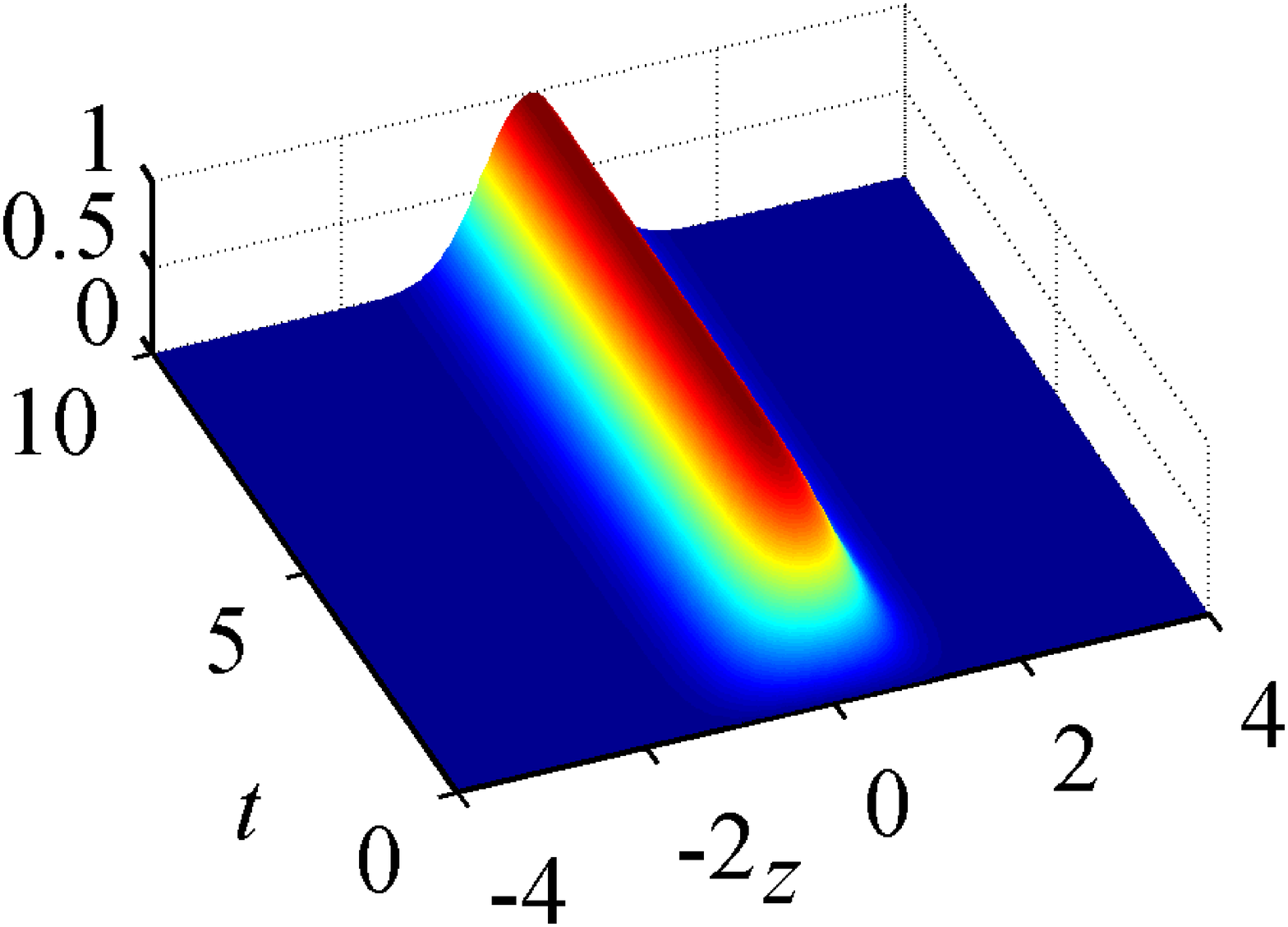}}\label{Fig:standa}
\subfigure[\ Thermal gas]
{\includegraphics[width=4cm]{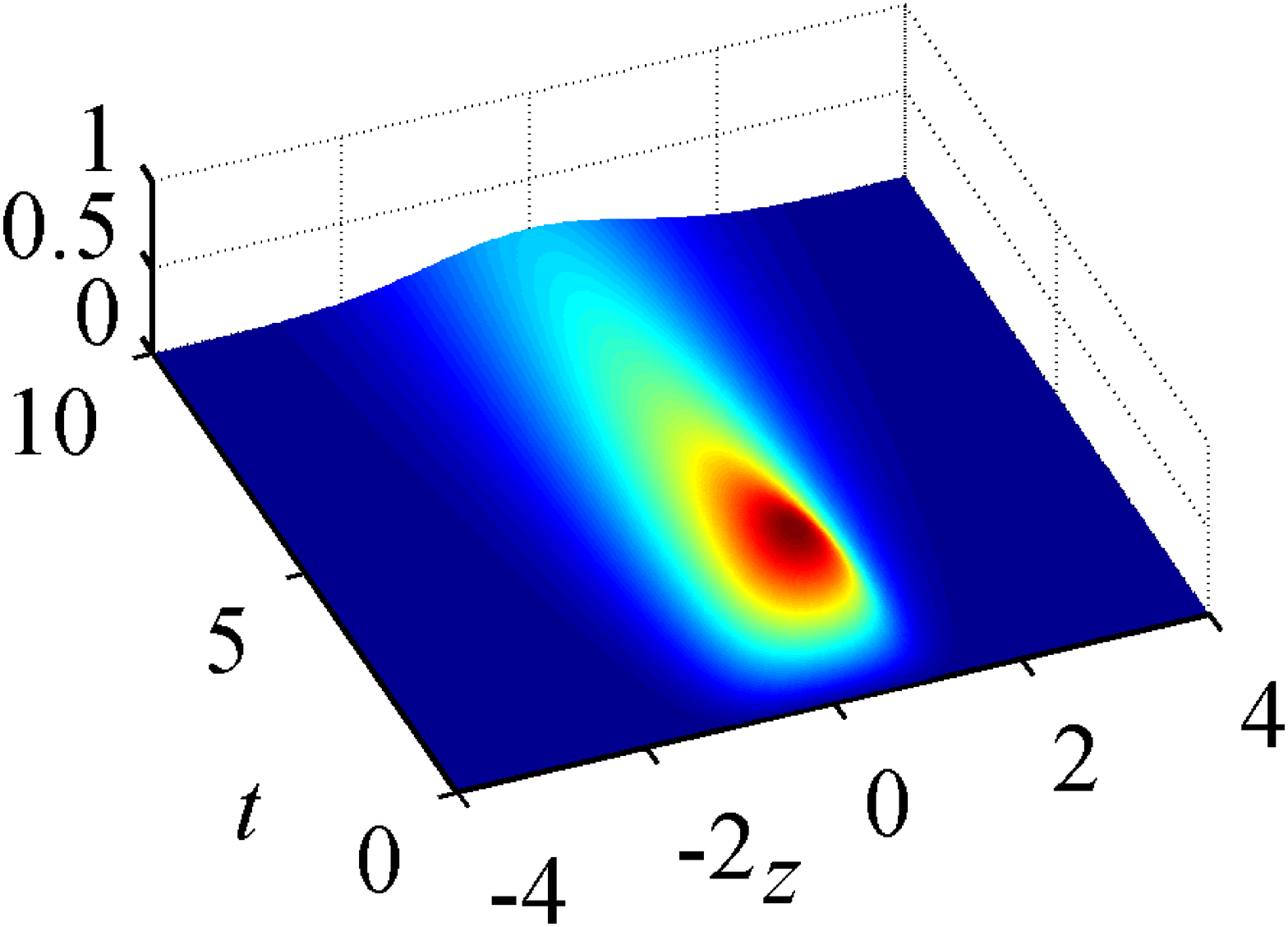}}\label{Fig:standb}
\caption{\label{Fig:stand}(color online) Retrieval of a stored probe
pulse with a standing wave coupling field. The probe field energy
density, in units of $\frac{\hbar\omega_p}{V}\abs{E_0}^2$, is
plotted for a medium comprised of stationary atoms (a) and thermal
atoms (b) as a function of $z$ in units of the pulse length $L_p$,
and $t$ in units of the switching time $T_s$. The absorption length
of the media is taken to be $l_a=0.1\times L_p$.}
\end{figure}

The components of the retrieved probe field found from
(\ref{Eq:def_polariton_stand}) are
\begin{subequations}
\begin{align}
E_p^+(z,t)&=\frac{1}{\sqrt{2}}\frac{\cos\theta(t)}{\cos\theta_0}
E_0\exp(-(z/L_p)^2)e^{-\Gamma_{bc}t}, \\
E_p^-(z,t)&=\frac{1}{\sqrt{2}}\frac{\cos\theta(t)}{\cos\theta_0}
E_0\exp(-(z/L_p)^2)e^{-\Gamma_{bc}t}.
\end{align}
\end{subequations}
In Fig.~\ref{Fig:stand} we compare the retrieval of an initially
stored probe pulse by a standing wave coupling field in the thermal
gas and ultra cold gas cases. The time dependence of the angle
$\theta$ is assumed to be given by
$\cos^2\theta(t)=\cos^2\theta_0\tanh(t/T_s)$ for $t\geq 0$, where
$T_s$ is the characteristic switching time. For simplicity, we have
assumed zero Raman dephasing $(\Gamma_{bc}=0)$ and taken the
characteristic length of the stored probe pulse to be $L_p=v_{g,0}
T_s$, where $v_{g,0}=c\cos^2\theta_0$. The probe field photon
density averaged over many wavelengths
$\abs{E_p^+}^2+\abs{E_p^-}^2$, in units of the photon density prior
to storage $\abs{E_0}^2$, is plotted as a function of $z$ in units
of $L_p$ and $t$ in units of $T_s$. In both the stationary atom case
and the thermal gas case we see that the stored probe pulse is
revived into a stationary probe field, but we note that in the
stationary atom case, the diffusive broadening of the probe field,
evident in the thermal gas case, is absent.

The solution for thermal gas media is based on the theory in
\cite{Zimmer}, which is reviewed briefly in appendix
\ref{Sec:Standing_wave_polaritons_thermal}. The medium is
characterized by the absorption length in the absence of EIT
$l_a=0.1\times L_p$, which roughly corresponds to the conditions in
\cite{Bajcsy}.

\subsection{Probe retrieval by a quasi-standing wave coupling field}
We shall now study the situation in which the probe field is
retrieved by a quasi-standing wave coupling field. In the previous
section, we saw that in the thermal gas case a quasi-standing wave
coupling field leads to a drift of the revived probe pulse in the
direction of the stronger of the two coupling field components.

In the ultra cold gas case considered here, we find from the
solution (\ref{Eq:DSP_stand_sol2}) that the revived probe pulse
instead splits into two parts. A stronger part which propagates in
the direction of the stronger of the coupling field components, and
a weaker part which propagates in the opposite direction.

\begin{figure}
\subfigure[\
$\Psi^+(z,t)$]{\includegraphics[width=4cm]{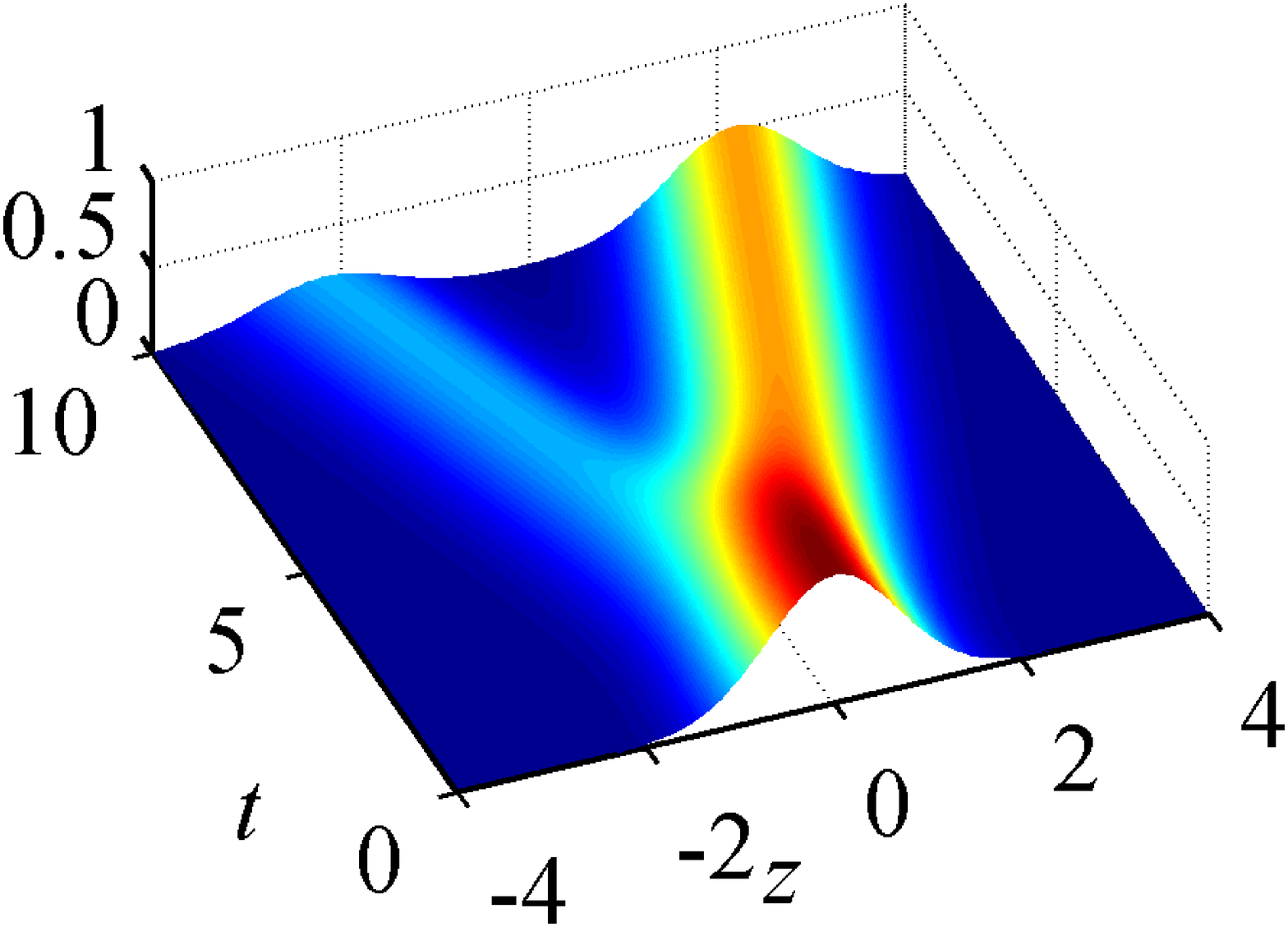}}
\label{Fig:qstand1a} \subfigure[\
$\Psi^-(z,t)$]{\includegraphics[width=4cm]{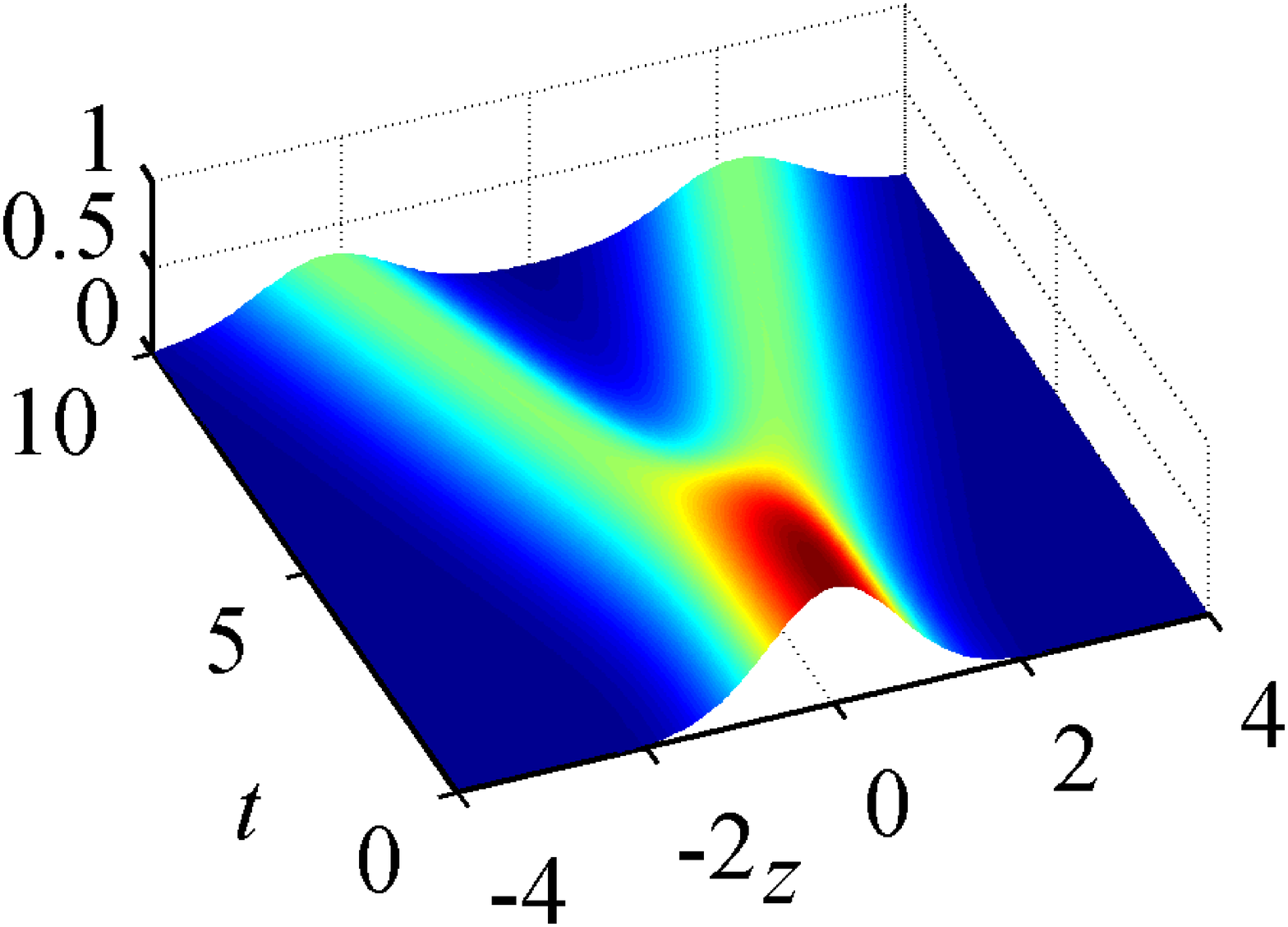}}
\label{Fig:qstand1b} \caption{\label{Fig:qstand1}Retrieval of a
stored probe pulse with a quasi-standing wave coupling field
($\kappa^+=\sqrt{0.55}$, $\kappa^-=\sqrt{0.45}$). Figures (a) and
(b) show the polariton amplitudes $\Psi^\pm$ in units of $\Psi_0$ as
a function of $z$ and $t$, in units of $L_p$ and $T_s$,
respectively.}
\end{figure}

\begin{figure}
\subfigure[\ Cold
gas]{\includegraphics[width=4cm]{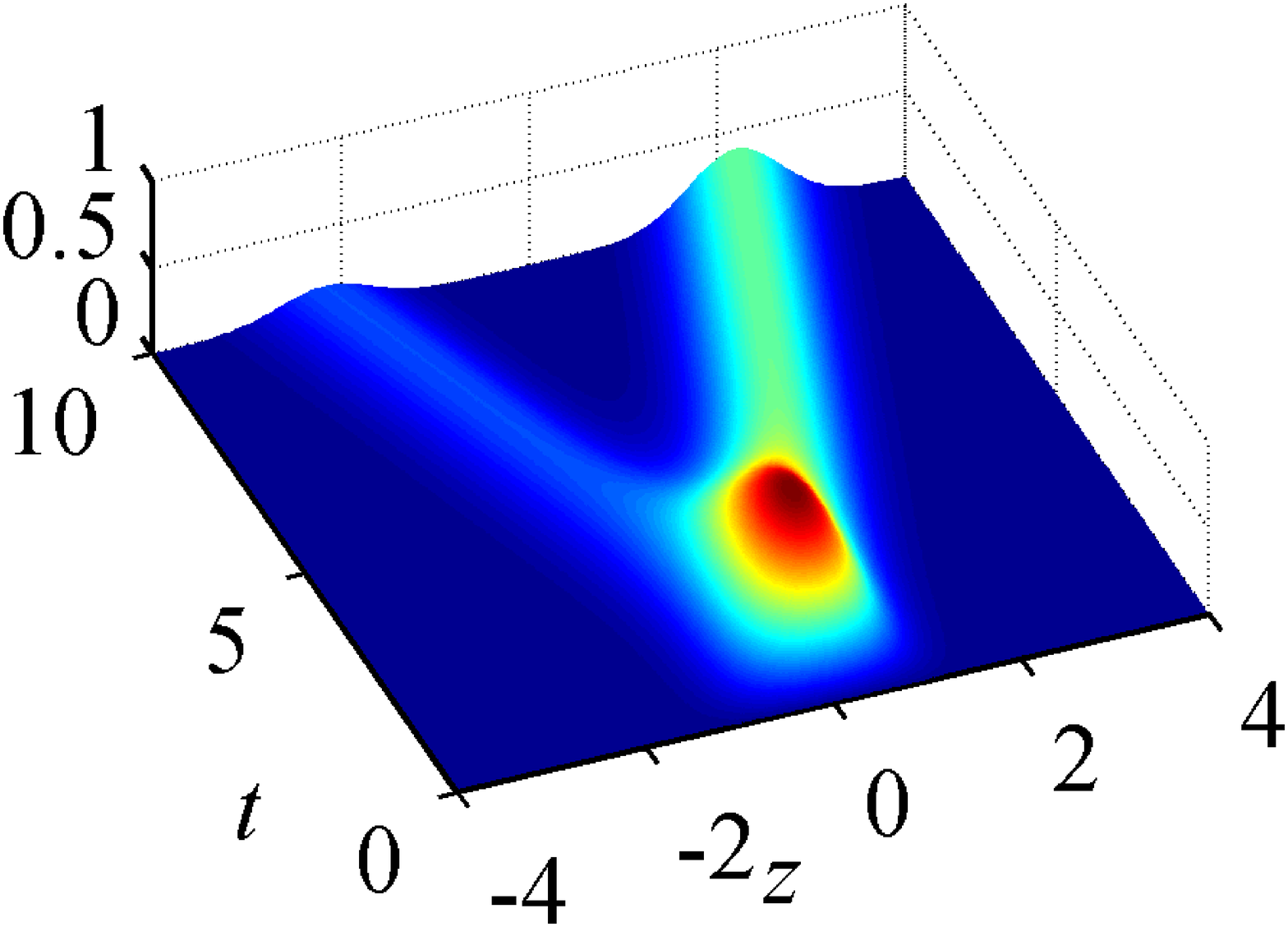}}\label{Fig:qstand2a}
\subfigure[\ Thermal
gas]{\includegraphics[width=4cm]{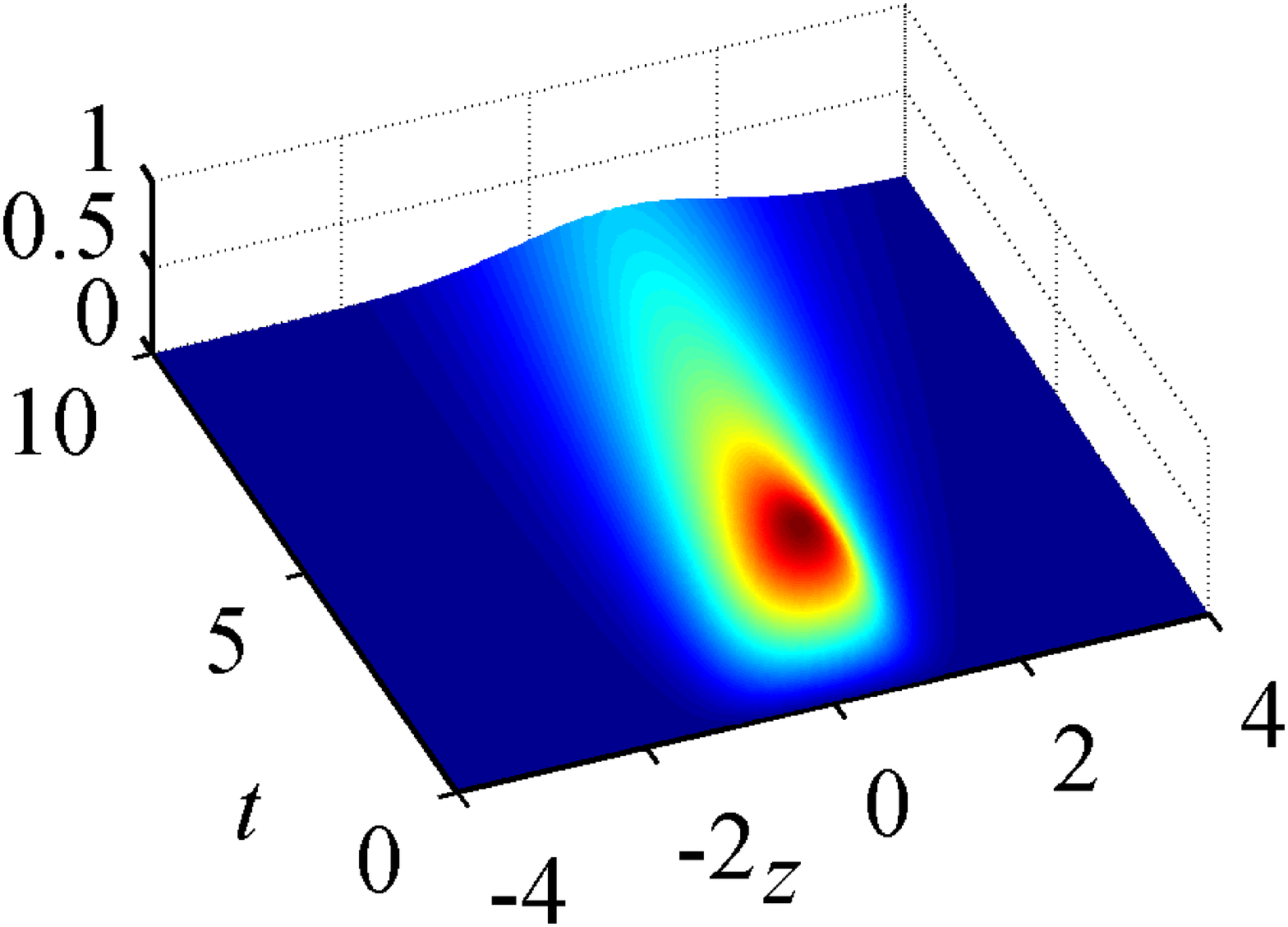}}\label{Fig:qstand2b}
\caption{\label{Fig:qstand2}Retrieval of a stored probe pulse with a
quasi-standing wave coupling field. The probe field energy density,
in units of $\frac{\hbar\omega_p}{V}\abs{E_0}^2$, is shown for both
the ultra cold gas case (a) and for the thermal gas case (b).
Parameters are the same as in Fig.~\ref{Fig:qstand1}.}
\end{figure}

Fig.~\ref{Fig:qstand1} shows the solution (\ref{Eq:DSP_stand_sol2})
with the same initial conditions as in Fig.~\ref{Fig:stand}, but
with $\kappa^+ =\sqrt{0.55}$ and $\kappa^-=\sqrt{0.45}$.
Fig.~\ref{Fig:qstand2} compares the retrieval of a stored probe
pulse by a quasi-standing wave coupling field in thermal and ultra
cold gas media. The splitting of the revived probe pulse is clearly
evident in the cold gas case, indicating a qualitative difference
between the thermal gas and the ultra cold gas cases. The cause of
this difference is the coupling to the high spatial-frequency
components of the Raman coherence $\hat{\sigma}_{bc}$ in the ultra
cold gas case. This splitting of the probe pulse is very important
when considering various schemes for interacting pulses. In the
phase-gate proposal of Andr\'e et al. \cite{Andre2}, a small
imbalance in the two components of the coupling field is used to
propagate a quasi-stationary light pulse across a stored excitation
in a thermal gas medium. As is evident from Fig.~\ref{Fig:qstand2}
this scheme would not work in media comprised of stationary atoms,
since a large part of the revived probe field would then propagate
in the wrong direction.

\subsection{Calculation of the Raman coherence}
To calculate the Raman coherence of the atoms, we use the zeroth
order expression (\ref{Eq:sigma_bc_stand_adia}) for $\sigma_{bc}$
\begin{equation}
\sqrt{N_\mathbf{r}}\sigma_{bc}=-\frac{g_p\sqrt{N_\mathbf{r}}
E_p}{\Omega_c}.
\end{equation}
By inserting the decompositions of the probe and coupling fields, as
well as the definition (\ref{Eq:def_polariton_stand}) of the DSP
field, we get
\begin{equation}
\sqrt{N_z}\sigma_{bc}=-\sin\theta\frac{\Psi^+(z,t)e^{ikz}+
\Psi^-(z,t)e^{-ikz}}{\kappa^+e^{ikz}+\kappa^-e^{-ikz}}.
\end{equation}
Inserting the solution (\ref{Eq:DSP_stand_sol2}) into this
expression, we find by a binomial expansion
\begin{equation}%\label{Eq:sigma_bc_cold2}
\begin{split}
\sqrt{N_z}\sigma_{bc}&=-\frac{1}{2}\sin\theta\biggl(\Psi(z-\beta
r,0)+\Psi(z+\beta r,0) \\
&\quad+\frac{\beta}{\abs{\kappa^+}^2}[\Psi(z-\beta r,0)-\Psi(z+\beta
r,0)] \\
&\quad\times\sum_{n=0}^\infty\left(-\frac{\kappa^-}{\kappa^+}
\right)^n e^{-2inkz}\biggr)e^{-\Gamma_{bc}t}.
\end{split}
\end{equation}
From this expression we see that the Raman coherence can be written
as
\begin{equation}\label{Eq:sigma_bc_expansion}
\sigma_{bc}(z,t)=\sum_{n=-\infty}^\infty\sigma_{bc}^{(2n)}(z,t)e^{2inkz},
\end{equation}
where the dc component is
\begin{equation}
\begin{split}
\sqrt{N_\mathbf{r}}\sigma_{bc}^{(0)}&=-\frac{1}{2}\sin\theta\biggl[
\biggl(1+\frac{\beta}{\abs{\kappa^+}^2}\biggr)\Psi(z-\beta r(t),0)\\
&+\biggl(1-\frac{\beta}{\abs{\kappa^+}^2}\biggr)\Psi(z+\beta r(t),0)
\biggr]e^{-\Gamma_{bc}t}.
\end{split}
\end{equation}
For the rapidly varying components of the Raman coherence we find
\begin{equation}
\begin{split}
\sqrt{N_\mathbf{r}}\sigma_{bc}^{(-2n)}&=-\frac{1}{2}\sin\theta
\frac{\beta}{\abs{\kappa^+}^2}\bigl(\Psi(z-\beta r(t),0) \\
&-\Psi(z+\beta r(t),0)\bigr)\left(-\frac{\kappa^-}{\kappa^+}
\right)^n e^{-\Gamma_{bc}t}
\end{split}
\end{equation}
and
\begin{equation}
\sqrt{N_\mathbf{r}}\sigma_{bc}^{(2n)}=0
\end{equation}
where, in both cases, $n>0$.

In the case of a perfect standing wave coupling field, only the dc
component of the Raman coherence is present which is given by
\begin{equation}
\sqrt{N_z}\sigma_{bc}^{(0)}(z,t)=-\sin\theta(t)\Psi(z,0)
e^{-\Gamma_{bc}t}.
\end{equation}
In the quasi-standing wave case, the rapidly varying components of
the Raman coherence $\sigma_{bc}^{(2n)}$ with \emph{negative} values
of $n$ attain a small but non-vanishing value, becoming
progressively smaller with decreasing $n$. The rapidly varying
components of the Raman coherence with \emph{positive} values of $n$
all vanish. An asymmetry in the Raman coherence is to be expected,
since neither the coupling field nor the revived probe field is
symmetric in $z$.

\section{Non-adiabatic corrections}\label{Sec:Non-adiabatic}
In \cite{Fleischhauer2} it was shown that the finite length of the
probe pulse leads to a broadening of the pulse envelope due to
dispersion. In this section we shall investigate the same effect in
the standing wave case and show that the dispersive broadening
vanishes in the case of a pure standing wave coupling field.

Our starting point is the differential equation
(\ref{Eq:diffeqn_sigmabc}) for the Raman coherence $\sigma_{bc}$. To
first order in $(\gamma_{ba}T)^{-1}$ we find
\begin{equation}
\sigma_{bc}=-\frac{g_pE_p}{\Omega_c}+
\frac{\Gamma_{ba}}{\abs{\Omega_c}^2}\ddt{}\left(
\frac{g_pE_p}{\Omega_c}\right),
\end{equation}
where we have assumed $\Gamma_{bc}=0$ to simplify the calculations.
Inserting this expression into (\ref{Eq:weak_probe_HL_stand_a}) and
introducing the DSP fields defined in
(\ref{Eq:def_polariton_stand}), we get
\begin{widetext}
\begin{equation}\label{Eq:sigma_ba_stand_nonadia}
\begin{split}
\sqrt{N_\mathbf{r}}\sigma_{ba}&=\frac{-\sin\theta}{i\Omega\bigl(1+
2\abs{\kappa^+}\abs{\kappa^-}\cos(2kz+\phi)\bigr)}\ddt{}\left(
\Psi^+e^{ikz}+\Psi^-e^{-ikz}\right) \\
&+\frac{\Gamma_{ba}}{g_p^2N_\mathbf{r}}
\frac{\sin\theta\tan^2\theta}{i\Omega\bigl(1+2\abs{\kappa^+}
\abs{\kappa^-}\cos(2kz+\phi)\bigr)^2}\dddtdt{}\left(\Psi^+e^{ikz}+
\Psi^-e^{-ikz}\right)
\end{split}
\end{equation}
\end{widetext}
where we have assumed that the coupling laser Rabi frequency changes
slowly enough to set $\dot{\theta}=0$ in the equations.

As in section \ref{Sec:Standing_wave_polaritons_cold} we need to
find the Fourier components $\sigma_{ba}^\pm$. To do this we apply
the Fourier series (\ref{Eq:Fourier_series1}) and we introduce
\begin{equation}
\frac{1}{(1+y\cos x)^2}=\frac{d_0}{2}+\sum_{n=1}^\infty d_n\cos(nx)
\end{equation}
where, as before, $y=2\abs{\kappa^+}\abs{\kappa^-}$ and
$x=2kz+\phi$. Inserting the Fourier series into
(\ref{Eq:sigma_ba_stand_nonadia}), we get
\begin{subequations}\label{Eq:sigma_ba_stand_nonadia2}
\begin{align}
\begin{split}
\sqrt{N_\mathbf{r}}\sigma_{ba}^+&=\frac{\sin\theta}{2i\Omega}\biggl(
-\ddt{}\left(a_0\Psi^++a_1e^{i\phi}\Psi^-\right) \\
&+\frac{\Gamma_{ba}}{g_p^2N_\mathbf{r}}\tan^2\theta\dddtdt{}\left(d_0
\Psi^++d_1e^{i\phi}\Psi^-\right)\biggr)
\end{split} \\
\begin{split}
\sqrt{N_\mathbf{r}}\sigma_{ba}^-&=\frac{\sin\theta}{2i\Omega}\biggl(
-\ddt{}\left(a_0\Psi^-+a_1e^{-i\phi}\Psi^+\right) \\
&+\frac{\Gamma_{ba}}{g_p^2N_\mathbf{r}}\tan^2\theta\dddtdt{}\left(d_0
\Psi^-+d_1e^{-i\phi}\Psi^+\right)\biggr)
\end{split}
\end{align}
\end{subequations}
The Fourier coefficients $a_{0,1}$ have already been calculated and
are given by (\ref{Eq:Fouriercoef_a}), while the Fourier
coefficients $d_{0,1}$ are given by
\begin{subequations}\label{Eq:Fouriercoef_d}
\begin{align}
d_0&=\frac{1}{\pi}\int_{-\pi}^\pi\frac{\mathrm{d}x}{(1+y\cos x)^2}
=\frac{2}{(1-y^2)^{3/2}}, \\
d_1&=\frac{1}{\pi}\int_{-\pi}^\pi\frac{\cos x\mathrm{d}x}{(1+y\cos
x)^2}=-\frac{2y}{(1-y^2)^{3/2}}.
\end{align}
\end{subequations}
We now insert the expressions (\ref{Eq:sigma_ba_stand_nonadia2})
into (\ref{Eq:probe_waveeqn}) to obtain a set of coupled wave
equations for the DSP fields
\begin{widetext}
\begin{subequations}
\begin{align}
\ddt{\Psi^+}+c\cos^2\theta'\ddz{\Psi^+}&=-\sin^2\theta'\biggl[
se^{i\phi}\ddt{\Psi^-}-\frac{\Gamma_{ba}}{g_p^2N_\mathbf{r}}
\tan^2\theta\dddtdt{}\left(
s'\Psi^++s''e^{i\phi}\Psi^-\right)\biggr] \\
\ddt{\Psi^-}-c\cos^2\theta'\ddz{\Psi^-}&=-\sin^2\theta'\biggl[
se^{-i\phi}\ddt{\Psi^+}-\frac{\Gamma_{ba}}{g_p^2N_\mathbf{r}}
\tan^2\theta\dddtdt{}\left(
s'\Psi^-+s''e^{-i\phi}\Psi^+\right)\biggr]
\end{align}
\end{subequations}
\end{widetext}
where we have introduced the constants
\begin{equation}
s=\frac{a_1}{a_0}\qquad s'=\frac{d_0}{a_0}\qquad
s''=\frac{d_1}{a_0}.
\end{equation}

Once again we consider the low group velocity limit $\cos^2\theta\ll
1$. With this approximation the wave equations simplify to
\begin{widetext}
\begin{subequations}\label{Eq:Polariton_waveeqn3}
\begin{align}
\ddt{\Psi^+}+c\cos^2\theta'\ddz{\Psi^+}&=-se^{i\phi}\ddt{\Psi^-}
+\frac{\Gamma_{ba}}{g_p^2N_\mathbf{r}}\tan^2\theta\dddtdt{}\left(
s'\Psi^++s''e^{i\phi}\Psi^-\right)\\
\ddt{\Psi^-}-c\cos^2\theta'\ddz{\Psi^-}&=-se^{-i\phi}\ddt{\Psi^+}
+\frac{\Gamma_{ba}}{g_p^2N_\mathbf{r}}\tan^2\theta\dddtdt{}\left(
s'\Psi^-+s''e^{-i\phi}\Psi^+\right)
\end{align}
\end{subequations}
\end{widetext}
To the same order of approximation, we can replace the second time
derivatives of the DSP fields with the second time derivative of the
zeroth order solution. Differentiating both sides of
(\ref{Eq:Polariton_waveeqn3}) with respect to $t$, and discarding
derivatives of order greater than two, we get
\begin{subequations}
\begin{align}
\dddtdt{\Psi^+}+c\cos^2\theta'\ddz{}\ddt{\Psi^+}&=-se^{i\phi}
\dddtdt{\Psi^-} \\
\dddtdt{\Psi^-}-c\cos^2\theta'\ddz{}\ddt{\Psi^-}&=-se^{-i\phi}
\dddtdt{\Psi^+},
\end{align}
\end{subequations}
where we once again assume that the coupling laser Rabi frequency
changes slowly. Using (\ref{Eq:Polariton_waveeqn3}) we can solve for
the second time derivatives of the DSP field. We find
\begin{equation}\label{Eq:Second_deriv_relation}
\dddtdt{\Psi^\pm}=\frac{1-y^2}{1-s^2}\left(c\cos^2\theta\right)^2
\dddzdz{\Psi^\pm},
\end{equation}
where we exploited the fact that in the low group velocity limit
$\cos^2\theta'\simeq\sqrt{1-y^2}\cos^2\theta$.
%\begin{equation}
%\cos^2\theta'\simeq\sqrt{1-y^2}\cos^2\theta.
%\end{equation}
Inserting (\ref{Eq:Second_deriv_relation}) into
(\ref{Eq:Polariton_waveeqn3}), and assuming that
$\abs{\kappa^+}\geq\abs{\kappa^-}$, the coupled wave equations take
the form
\begin{widetext}
\begin{subequations}\label{Eq:Polariton_waveeqn4}
\begin{align}
\ddt{\Psi^+}+\abs{\kappa^+}^2v_g\ddz{\Psi^+}&=\kappa^+\kappa^{-*}v_g
\ddz{\Psi^-}+\frac{\abs{\kappa^+}^2l_av_g}{\sqrt{1-y^2}}\dddzdz{}
\left(\abs{\kappa^+}^2\Psi^+-\kappa^+\kappa^{-*}\Psi^-\right) \\
\ddt{\Psi^-}-\abs{\kappa^+}^2v_g\ddz{\Psi^-}&=-\kappa^{+*}\kappa^-v_g
\ddz{\Psi^+}+\frac{\abs{\kappa^+}^2l_av_g}{\sqrt{1-y^2}}\dddzdz{}
\left(\abs{\kappa^+}^2\Psi^--\kappa^{+*}\kappa^-\Psi^+\right)
\end{align}
\end{subequations}
\end{widetext}
To solve the coupled wave equations (\ref{Eq:Polariton_waveeqn4}) we
proceed by Fourier transforming with respect to $z$, such that
$\Psi^\pm(z,t)\to\tilde{\Psi}^\pm(q,t)$, and find the solution
\begin{widetext}
\begin{subequations}
\begin{align}
\begin{split}
\tilde{\Psi}^+(q,t)&=\frac{1}{2d}\biggl(\bigl(b\tilde{\Psi}^-(q,0)-
(\abs{\kappa^+}^2-d)\tilde{\Psi}^+(q,0)\bigr)\exp\bigl(iq\lambda_+
r(t)\bigr) \\
&+\bigl( (\abs{\kappa^+}^2+d)\tilde{\Psi}^+(q,0)-b\tilde{\Psi}^-
(q,0)\bigr)\exp\bigl(iq\lambda_-r(t)\bigr)\biggr)
\end{split} \\
\begin{split}
\tilde{\Psi}^-(q,t)&=\frac{1}{2d}\biggl(\bigl(-b^*\tilde{\Psi}^+(q,0)
+(\abs{\kappa^+}^2+d)\tilde{\Psi}^-(q,0)\bigr)\exp\bigl(iq\lambda_+
r(t)\bigr) \\
&+\bigl(b^*\tilde{\Psi}^+(q,0)-(\abs{\kappa^+}^2-d)\tilde{\Psi}^-
(q,0)\bigr)\exp\bigl(iq\lambda_-r(t)\bigr)\biggr)
\end{split}
\end{align}
\end{subequations}
\end{widetext}
where
\begin{equation}
b=\kappa^+\kappa^{-*}(1-iq\xi),\quad
\lambda^\pm=i\abs{\kappa^+}^2\xi q\pm d
\end{equation}
and
\begin{equation}
\xi=\frac{\abs{\kappa^+}^2l_a}{\sqrt{1-y^2}},
\end{equation}
\begin{equation}
d=\sqrt{\abs{\kappa^+}^2\left(\abs{\kappa^+}^2-\abs{\kappa^-}^2\right)
-\abs{\kappa^+}^2\abs{\kappa^-}^2\xi^2q^2}.
\end{equation}

Inserting the initial conditions (\ref{Eq:Psi_pm_init_cond}) for the
DSP field and considering the limit $\kappa^\pm\rightarrow
\tfrac{1}{\sqrt{2}}$, corresponding to a pure standing wave coupling
field, the solution becomes
\begin{subequations}
\begin{align}
\Psi^+(z,t)&=\frac{1}{\sqrt{2}}\Psi(z,0), \\
\Psi^-(z,t)&=\frac{1}{\sqrt{2}}\Psi(z,0).
\end{align}
\end{subequations}
From this solution it is clear that the broadening of the pulse
envelope due to dispersion is absent in the case of a pure standing
wave coupling field. The effect \emph{is} present in the case of a
quasi-standing wave coupling field. If we consider the limiting case
of a traveling wave coupling field $(\kappa^-\rightarrow 0)$, we
find the same dispersion term in the wave equation
(\ref{Eq:Polariton_waveeqn4}) that is given in \cite{Fleischhauer2}.

\section{Summary}\label{Sec:Summary}
In this article we have presented a detailed theoretical treatment
of stationary light pulses in media comprised of stationary atoms,
such as ultra cold gasses and solid state media. We found that
contrary to the thermal gas case, the achievable trapping time is
limited only by the Raman dephasing rate of the atoms and such media
are thus ideally suited for the kind of nonlinear optical
interactions envisaged in \cite{Friedler,Andre2}. It was also shown
that the behavior of the probe pulse when employing quasi-stationary
coupling fields is significantly different for moving and non-moving
atoms. This fact must be taken into account when considering schemes
for interacting pulses. Although, to the best of our knowledge, no
experiment with stationary light pulses in ultra cold media has yet
been reported, several experiments on normal EIT and light storage
have been performed with ultra cold gasses \cite{Liu} and solid
state media \cite{Turukhin,Longdell}. These experiments have also
demonstrated the possibility of using beam geometries other than
copropagating probe and coupling lasers. We therefore expect that
the experimental demonstration of stationary light pulses in such
media is within present day capability.

\begin{acknowledgments}
We gratefully acknowledge stimulating discussions with M.
Fleischhauer and F. Zimmer, and we thank A. Andr\'e and M. Lukin for
communicating their results on stationary pulses in thermal gasses
prior to publication in \cite{Zimmer}. This work is supported by the
European Integrated Project SCALA and the ONR-MURI collaboration on
quantum metrology with atomic systems.
\end{acknowledgments}

\appendix
\section{Standing wave polaritons in thermal gasses}
\label{Sec:Standing_wave_polaritons_thermal}

As shown in Sec.~\ref{Sec:Standing_wave_polaritons_cold} the
behavior of the stationary light pulses depends critically on
whether the EIT medium is comprised of stationary or moving atoms.
In this section we present a brief review of the theory for the
thermal gas case presented in \cite{Zimmer}. It is argued that the
motion of the atoms in a thermal gas causes a rapid dephasing of the
spatially rapidly varying components of the Raman coherence and it
is therefore assumed that only the $n=0$ component in the expansion
(\ref{Eq:sigma_bc_expansion}) is non-vanishing. Consequently, the
only non-vanishing components of the optical coherence in the
expansion (\ref{Eq:sigma_ba_expansion}) is the $n=-1$ and $n=0$
terms. With this approximation, the relevant Heisenberg-Langevin
equations for the slowly varying operators are
\begin{subequations}\label{Eq:HL_thermal}
\begin{align}
\dot{\sigma}_{ba}^+&=ig_pE_p^++i\Omega_c^+\sigma_{bc}-\Gamma_{ba}
\sigma_{ba}^+, \\
\dot{\sigma}_{ba}^-&=ig_pE_p^-+i\Omega_c^-\sigma_{bc}-\Gamma_{ba}
\sigma_{ba}^-, \\
\dot{\sigma}_{bc}&=i(\Omega_c^{+*}\sigma_{ba}^++\Omega_c^{-*}
\sigma_{ba}^-)-\Gamma_{bc}\sigma_{bc}.\label{Eq:HL_thermal_c}
\end{align}
\end{subequations}
As shown in \cite{Zimmer} these equations can be solved
approximately by adiabatically eliminating the optical coherences
$\sigma_{ba}^\pm$ and making an adiabatic expansion of
(\ref{Eq:HL_thermal_c}). The resulting expressions for the
components of the optical coherence $\sigma_{ba}^\pm$ is then
inserted into the wave equations (\ref{Eq:probe_waveeqn}). Contrary
to \cite{Zimmer}, which deals directly with the probe field
operators $E_p^\pm$, we introduce the polariton field defined by
(\ref{Eq:def_polariton_stand}) which enables us to treat time
dependent coupling fields in a consistent manner. To facilitate the
solution of the resulting wave equations, sum and difference normal
modes defined by
\begin{subequations}
\begin{align}
\Psi_S&=\kappa^{+*}\Psi^++\kappa^{-*}\Psi^-, \\
\Psi_D&=\kappa^-\Psi^+-\kappa^+\Psi^-
\end{align}
\end{subequations}
are introduced. In the case of an optically thick medium, the
difference mode can be adiabatically eliminated, resulting in a
diffusion equation for the sum normal mode
\begin{widetext}
\begin{equation}\label{Eq:PsiS_waveeqn}
\ddt{\Psi_S}+(\abs{\kappa^+}^2-\abs{\kappa^-}^2)c\cos^2\theta
\ddz{\Psi_S}=4\abs{\kappa^+}^2\abs{\kappa^-}^2l_a c\cos^2\theta
\dddzdz{\Psi_S}-\Gamma_{bc}\sin^2\theta\Psi_S,
\end{equation}
\end{widetext}
where we have assumed zero probe field detuning $(\delta_p=0)$. The
difference normal mode is given by
\begin{equation}\label{Eq:PsiD_waveeqn}
\Psi_D=-2\kappa^+\kappa^-l_a\ddz{\Psi_S}.
\end{equation}
The solution of (\ref{Eq:PsiS_waveeqn}), subject to the initial
conditions (\ref{Eq:Psi_pm_init_cond}), is the basis for the
comparison between thermal gas media and stationary atom media
presented in Sec.~\ref{Sec:Standing_wave_polaritons_cold}.

%%% Bibliography %%%

\end{document}